\documentclass[]{iopart}
\usepackage{iopams}
\usepackage[english]{babel}
\usepackage[dvips]{graphicx}
\eqnobysec
\begin{document}
\newcommand{\beq}{\begin{eqnarray}}
\newcommand{\eeq}{\end{eqnarray}}
\newcommand{\ud}{\mathrm{d}}
\newcommand{\udd}{\frac{\mathrm{d}}{\mathrm{d}t}}
\newcommand{\lt}{\left(}
\newcommand{\rt}{\right)}
\newcommand{\lqu}{\left[}
\newcommand{\rqu}{\right]}
\newcommand{\vp}{\varphi}
\newcommand{\zint}{Z_{\textsc{int}}(\varphi)}
\newcommand{\zintop}{\hat{Z}_{\textsc{int}}(\hat{\varphi})}
\newcommand{\dota}{\dot{a}}
\newcommand{\ddota}{\ddot{a}}
\newcommand{\dotp}{\dot{\phi}}
\newcommand{\ddotp}{\ddot{\phi}}
\newcommand{\dotvp}{\dot{\varphi}}
\newcommand{\ddotvp}{\ddot{\varphi}}
\newcommand{\lag}{\mathcal{L}}
\newcommand{\vxp}{v - \xi \phi^2}
\newcommand{\vxpop}{v - \xi \hat{\phi}^2}
\newcommand{\wx}{W_\xi (\phi)}
\newcommand{\wxop}{\hat{W}_\xi (\hat{\phi})}
\newcommand{\wxopm}{\lan \hat{W}_\xi (\hat{\phi}) \ran}
\newcommand{\lan}{\left<}
\newcommand{\ran}{\right>}
\newcommand{\pop}{\hat{\phi}}
\newcommand{\vpop}{\hat{\varphi}}
\newcommand{\pa}{P_a}
\newcommand{\paop}{\hat{P}_a}
\newcommand{\pp}{P_\phi}
\newcommand{\ppop}{\hat{P}_\phi}
\newcommand{\pvp}{P_\varphi}
\newcommand{\pvpop}{\hat{P}_\varphi}
\newcommand{\ptau}{\partial_\tau}
\newcommand{\hs}{\hat{H}_\textsc{s}}
\newcommand{\lp}{\ell_{\rm P}}
\newcommand{\hub}{\mathcal{H}}
\newcommand{\ct}{\tilde{\chi}}
\newcommand{\hc}{\hat{H}_{\chi,0}}
\renewcommand{\bi}{\hat{b}}
\newcommand{\bid}{\hat{b}^\dagger}
\newcommand{\chis}{\left. | \chi_s \ran}
\newcommand{\para}{\partial_a}
\newcommand{\kck}{\ket{\chi_k}}
\newcommand{\bra}[1]{\mbox{$\langle #1\! \mid$}}
\newcommand{\bbra}[1]{\mbox{$\left\langle #1 \right\mid$}}
\newcommand{\ket}[1]{\mbox{$\mid \!#1\rangle$}}
\newcommand{\bket}[1]{\mbox{$\left\mid #1\right\rangle$}}
\newcommand{\pro}[2]{\mbox{$\langle #1 \mid #2\rangle$}}
\newcommand{\vm}[1]{\mbox{$\langle #1\rangle$}}
\newcommand{\hh}{\hat{H}}
\newcommand{\hd}{\hat{\Delta}}
\newcommand{\kc}{\ket{\chi}}
\newcommand{\bc}{\bra{\chi}}
\newcommand{\kcs}{\ket{\chi_{\rm s}}}
\title{Hints of (trans-Planckian) asymptotic freedom in semiclassical cosmology}
\author{Corrado Appignani and Roberto Casadio}
\address{Dipartimento di Fisica, Universit\`a di Bologna,
and I.N.F.N., Sezione di Bologna,
via Irnerio 46, 40126 Bologna, Italy.}
\eads{\mailto{appignani@bo.infn.it},
\mailto{casadio@bo.infn.it}}
\begin{abstract}
We employ the semiclassical approximation to the
Wheeler-DeWitt equation in the spatially flat de~Sitter Universe
to investigate the dynamics of a minimally coupled
scalar field near the Planck scale.
We find that, contrary to na\"ive intuition, the effects of
quantum gravitational fluctuations become negligible
and the scalar field states asymptotically
approach plane-waves at very early times.
These states can then be used as initial conditions for the quantum
states of matter to show that each mode essentially originated
in the minimum energy vacuum.
Although the full quantum dynamics cannot be solved exactly
for the case at hand, our results can be considered as supporting
the general idea of asymptotic safety in quantum gravity.
\end{abstract}
\section{Introduction}
A proper treatment of any matter-gravity system would require
the quantisation of all degrees of freedom of matter and gravity.
Attempts in this direction include String Theory~\cite{ST}
and Loop Quantum Gravity~\cite{LQG}.
While awaiting for a universally accepted theory of quantum
gravity, many useful results have been hitherto obtained by quantising the
matter fields on a classical background~\cite{birrell}.
The limit of this approach can be na\"ively identified with the requirement
that only energies (lengths) below (above) the Planck scale
should be (explicitly) involved in the description of physical
processes.
There are however notorious exceptions to this rule, namely the
occurrence of trans-Planckian frequencies in the Hawking
effect~\cite{hawking} and in inflationary models~\cite{branden},
the latter being the case we shall consider hereafter.
\par
According to the above remark, in a Friedman-Robertson-Walker Universe,
physical wavelengths of matter fields should be allowed only if $a/k\gtrsim\lp$
($k$ being the wavenumber) or, equivalently, the cosmic scale factor $a$ may
be treated as a classical background quantity only when it is
larger than multiples of the Planck length $k\,\lp$ for the modes
$k$ of interest.
Such a condition is easily violated during the inflationary evolution,
since modes that are detected in the CMB spectrum today 
were trans-Planckian at the beginning of inflation~\cite{branden}.
Nonetheless, the results obtained from quantum field theory
are in very good agreement with observation.
\par
Many papers can be found in the literature which try to explain
why such an approximate method works so well by showing that possible
corrections to the CMB spectrum due to the unknown trans-Planckian
physics should be small~\cite{dani,tp,lim}. 
For example, in Ref.~\cite{dani}, it was suggested that the trans-Planckian
phase of the evolution results into suitable initial conditions for the quantum
matter states at the time when the corresponding field modes become
sub-Planckian and can afterwards be analysed by the usual means.
Several proposals for the initial conditions were also put forward based on the
principle of minimum uncertainty~\cite{dani} or otherwise~\cite{lim}.
There are two main critiques to this viewpoint:
since different modes become sub-Planckian at
different times, one must assume that the background metric
is already classical {\em before\/} any matter modes become
sub-Planckian (tantamount to requiring that quantum
gravitational fluctuations be small);
moreover, any choice of initial conditions appears arbitrary
without a more explicit knowledge of the trans-Planckian theory.
Although several approaches support the possibility that the Universe
was indeed born in the semiclassical regime
(with a non-zero initial scale factor~\cite{hs,loll,gf}),
the issue of initial conditions for the matter states remains open.
\par
We shall here investigate the dynamics of matter fields in cosmology
near the limit of validity of the semiclassical approximation,
that is for $a\sim k\,\lp$ when the proper frequency $\omega\sim k/a$
of matter modes approaches the Planck scale.
In order to perform our analysis, we shall employ the semiclassical
approximation to the quantum Wheeler-DeWitt equation~\cite{WDW}
in minisuperspace as proposed, for example, in Ref.~\cite{bv} and later developed in
Refs.~\cite{fvv,acg,xi0}  (for a recent review of the Wheeler-DeWitt
equation in cosmology, see Ref.~\cite{kiefer}).
In particular, our starting point will be a modified Schr\"odinger equation
for the quantum state of a mode of a minimally coupled scalar field
in the de~Sitter Universe as obtained within this approach in Ref.~\cite{xi0}.
This equation contains terms of order $\lp^2$ which were treated
as a perturbation at large times ({\em i.e.}, for large scale factor $a/k\gg\lp$),
and the corresponding corrections to the dispersion relation
and CMB power spectrum estimated.
We shall show in the next Section that, in the opposite regime
of very early times, quantum gravitational fluctuations become
negligible and the scalar field is asymptotically free at higher
and higher energies.
We shall also show that the initial state of each mode is very close to the
usual vacuum and the coupling to gravitational perturbations
always remains small.
Of course, Planckian physics is expected to be non-perturbative due to
the highly non-linear nature of the Einstein-Hilbert action.
Since the Wheeler-DeWitt equation cannot be solved exactly for the case
at hand, we cannot give a rigourous mathematical proof that our results are
totally safe.
However, asymptotic safety in quantum gravity was conjectured almost
thirty years ago~\cite{weinberg} and seems to be further supported
now~\cite{souma} in the effective action approach of quantum field theory.
We thus believe that our results can be considered as also supporting this
general idea from a totally different perspective.
\par
We shall use units with $c=\hbar=1$.
\section{Scalar field evolution at very early times}
\label{sec:main}
In Ref.~\cite{xi0}, the evolution of one mode $\phi=\phi^{(k)}$
of a scalar field minimally coupled to gravity was studied by means of the
Born-Oppenheimer reduction of the Wheeler-deWitt
equation in the minisuperspace of $a$ and $\phi$~\cite{bv}.
We shall not repeat the derivation here, but just recall that
using this procedure, one obtains the semiclassical
Einstein-Hamilton-Jacobi equation for $a=a(\tau)$ and a perturbed
Schr\"odinger equation for $\kc$, the quantum state of $\phi$.
Both equations contain terms proportional to $\lp^2$, which can be
viewed as representing quantum gravitational perturbations, but we shall
primarily consider here their effects in the matter equation.
\par
The case of interest to us is thus represented by the equation for $\kc$ in
the de~Sitter space-time,
\beq
a=a_0\,e^{\hub\,\tau}
\ .
\label{ds}
\eeq
where $\hub$ is the Hubble constant.
We shall also restrict our analysis to scalar field modes with wavelength much
shorter than the Hubble length,
\beq
a/k\ll\hub^{-1}
\ ,
\label{K}
\eeq
so that  the spatial curvature does not affect the scalar field
dynamics and can be neglected.
The relevant Schr\"odinger equation can then be written as (see equation~(18) in
Ref.~\cite{xi0})
\beq
\lt 1-\frac{3\,i\,\lp^2}{2\,a^3\,\hub} \rt
\lt i\,\ptau - \hh \rt \kc =
\frac{\lp^2}{2\,a^3\,\hub}\,\Delta\hat O \kc
\ ,
\label{start}
\eeq
where
\beq
\hh =
\frac{1}{2}\left(
\frac{\hat{\pi}^2}{a^3} +
a\,k^2\,\hat{\phi}^2
\right)
\label{H0}
\eeq
is the ``unperturbed'' Hamiltonian for the quantum state of the
mode $k$,
\beq
\hat{O} =
\frac{1}{\hub}\, \ptau^2
+\frac{2\,i}{\hub}\, \vm{\hh}\, \ptau
+3\,i\,\hh
\ ,
\label{def}
\eeq
and $\Delta\hat X \equiv \hat{X} - \vm{\hat{X}}$ for any operator $\hat X$, with
the scalar product between matter states given by the Schr\"odinger
product at fixed $a$, 
\beq
\langle\chi|\xi\rangle=\int \chi^*(\phi;a)\,\xi(\phi;a)\,\ud \phi
\ .
\label{scalprod}
\eeq
At zero order in $\lp$, the above equation describes the usual
quantum evolution in the cosmic time $\tau$ of a scalar field in the de~Sitter
space-time, including particle production induced by the expanding background
on the initial vacuum.
Terms of order $\lp^2$ therefore represent (leading-order)
effects of the quantum gravitational fluctuations on the
evolution of the scalar field.
By contracting~\eref{start} with $\bra{\chi}$,
we obtain
\beq
i \,\vm{\ptau} = \vm{\hh}
\ .
\label{aux}
\eeq
Note that this equation alone does not imply that the solutions to
equation~\eref{start} are eigenstates of the Hamiltonian $\hat H$.
In fact,~\eref{aux} follows from $\langle\Delta\hat O\rangle=0$
which means that $\Delta\hat O$ maps each state into orthogonal states.
\par
To make contact with observation, it is possible to define an effective
Hamiltonian which takes order $\lp^2$ effects into account and
consequently derive a modified dispersion relation and the
CMB power spectrum~\cite{xi0}. 
In Ref.~\cite{xi0}, all of these quantities were obtained from~\eref{start}
by investigating the late-time evolution of the system.
This was termed regime~II in that paper, its precise definition being
that the cosmic scale factor be so large that $a^3\,\hub \gg \lp^2$.
It was also stated that, for earlier times such that 
$a^3\, \hub \ll \lp^2$ (regime~I), the effect of gravitational fluctuations was
expected to be negligible.
To support this assertion, it was shown that the ratio between
the perturbations and the Hamiltonian vanishes in regime~I.
However, perturbative relations, strictly valid
only in regime~II, were used in that argument, 
which might appear questionable,
since one na\"ively expects that the strength of
quantum gravitational fluctuations grow indefinitely for increasing
mode energy ({\em i.e.}, going backward in time).
It is therefore necessary to further investigate the regime~I.
\subsection{Asymptotic states}
\label{sec:asy}
We start from~\eref{start}, which is valid in any regime
and, for convenience, we rewrite as
\beq
\lt i\,\ptau - \hh \rt \kc =
\frac{y}{1-3\,i\,y}\, \Delta\hat O \kc
\ ,
\label{start2}
\eeq
where 
\beq
y \equiv \frac{\lp^2}{2\,a^3\,\hub}
\ ,
\eeq
and the very early regime~I is then defined as $y \gg 1$ (or $-\tau\gg \hub^{-1}$).
\par
It is important to remark that the limit $y\gg 1$ is compatible with  the
semiclassical approximation from which~\eref{start} follows.
In fact, from $a/\lp \sim k$ and $y\gg 1$, one obtains the
compatibility condition $a\,\hub \ll k^{-2}$.
Further, upon requiring that the mode $k$ initially lies well inside the Hubble radius
(that is, condition~\eref{K}), one finally obtains
\beq
a\,\hub\ll k\ll {1}/{\sqrt{a\,\hub}}
\ ,
\label{cc1}
\eeq
which can be satisfied when $a\,\hub\ll 1$, a condition which is obviously
compatible with the spatially flat de~Sitter space-time~\eref{ds}
at very early times.
On the other hand, if the geometry of the early Universe was exactly that of a spatially
closed de~Sitter~\cite{hs}, $a\,\mathcal{H}\ge 1$ and no mode $k$ exists which
satisfies~\eref{cc1}.
In this case, however, $y\lesssim \ell_{\rm P}^2\,\mathcal{H}^2\ll 1$ and scalar field
modes are generated in the perturbative regime~II in which quantum gravitational
fluctuations were already shown to be small~\cite{xi0}.
\par
We can now trade $a=a(\tau)$ and, given~\eref{ds},
the time $\tau$ for $y$, so that
\beq
\partial_\tau=-3\,\hub\,y\,\partial_y
\ ,
\eeq
and then expand all the quantities in equation~\eref{start2} in powers of $y$.
For example, the operator representing gravitational fluctuations
becomes
\beq
\hat O=3\,\hub\,y\lqu
3\,\partial_y+3\,y\,\partial^2_y
-2\,i\,y\,\frac{\vm{\hat\pi^2}}{\lp^2}\,\partial_y
+i\,\frac{\hat\pi^2}{\lp^2}\rqu
\ ,
\eeq
and the asymptotic form of the scalar field Hamiltonian is given by
\beq
\hh=\hub\,y\lqu
\frac{\hat\pi^2}{\lp^2}
+\frac{k^2}{2}\lt\frac{\lp^2}{2\,\hub^4\,y^4}\rt^{1/3}\hat\phi^2\rqu
\simeq
\hub\,y\,\frac{\hat\pi^2}{\lp^2}
\ ,
\eeq
in which we neglected the potential term since it is proportional to $y^{-1/3}$.
After some trivial simplifications, equation~\eref{start2} takes on the asymptotic
(and dimensionless) form
\beq
\fl
\lt i\,\partial_y+\frac{\hat\pi^2}{3\,\lp^2}\rt\kc
\simeq
\Delta\lqu 
\frac{\hat\pi^2}{3\,\lp^2}
- 2\,\frac{\vm{\hat\pi^2}}{3\,\lp^2}\,y\,\partial_y
-i \lt 1 +y\,\partial_y\rt \partial_y\rqu
\kc
\ ,
\label{asyEq}
\eeq
which can be easily solved by writing
\beq
\kc=e^{i\,\varphi(y)}\,\ket{\kappa}
\ ,
\eeq
where $\ket{\kappa}$ are orthogonal eigenstates of the scalar
field momentum,
\beq
\frac{\hat\pi}{\sqrt{3}\,\lp}\ket{\kappa}=\kappa\,\ket{\kappa}
\ ,
\eeq
and do not depend on $y$.
The right end side of equation~\eref{asyEq} therefore vanishes on such states
and one finds
\beq
\varphi=\kappa^2\,y
\ ,
\eeq
so that the asymptotic ($-\tau=-\tau_{\rm asy}\gg \hub^{-1}$) un-normalised
quantum states of the scalar field mode $k$ can be finally written as
\beq
\fl
\ket{\chi_{\rm asy}}\sim
\exp\lt i\,\frac{\lp^2\,\kappa^2}{2\,a^3\,\hub}\rt
\ket{\kappa}
=\exp\lt -i\,\int\limits^{\tau_{\rm asy}}\frac{3\,\lp^2\,\kappa^2}{2\,a^3}\, \ud t\rt
\,
\sqrt{\frac{\sqrt{3}\,\lp}{2\,\pi}}
\,e^{\strut\displaystyle{i\,\sqrt{3}\,\kappa\,\lp\,\phi}}
\ ,
\label{sol}
\eeq
where the factor of $\sqrt{\lp}$ was included to make the scalar
product~\eref{scalprod} dimensionless, since $\phi$ is the inverse of a length.
Finally, we note that these solutions~\eref{sol} to the matter equation~\eref{start}
in the asymptotic regime~I  also satisfy the unperturbed Schr\"odinger equation 
\beq
i\,\partial_\tau\ket{\chi_{\rm asy}}
=
\hat H_{\rm asy}\,\ket{\chi_{\rm asy}}
\ ,
\eeq
with the free particle Hamiltonian
\beq
\hat H_{\rm asy}=\frac{\hat\pi^2}{2\,a^3}
\simeq
\frac{\hat\pi^2}{2\,\lp^3\,k^3}
\ .
\label{Hasy}
\eeq
Consequently, for very early times (when $a\sim k\,\lp$), the influence of
gravitational fluctuations does not grow indefinitely, but vanishes asymptotically.
\par
Technically speaking, the above plane-waves~\eref{sol} are not normalisable for
$\phi$ and $\kappa\in\mathbb{R}$ and their physical interpretation remains
somewhat unclear so far, also because of their dependence on the parameter
$\kappa$.
Further, it is easily seen that they do not solve equation~\eref{start} for finite values
of $y$ since, in this case, the potential in $\hat H$ is not negligible and the Hamiltonian
changes (approximately) into that of a harmonic oscillator (free scalar field) with
time-dependent frequency (and mass) which admits (approximate) normalizable
(discrete) eigenstates.
Thus, as soon as the potential is no more neglibile, the Hamiltonian spectrum
changes and particle production begins (moreover, $\Delta\hat O\kc\not=0$,
and transitions also occur among invariant eigenstates~\footnote{For more details about
invariant operators and states, see Refs.~\cite{lewis,acg,xi0}.
\label{finv}}).
The only way one may accommodate for the asymptotic states~\eref{sol} in the theory is
therefore by considering normalised wave-packets as initial conditions for the subsequent
evolution of the scalar field states~\footnote{This interpretation partly overcomes one of the
two criticisms to the approach of Ref.~\cite{dani}.}.
We shall however see that one can still draw significant conclusions without dealing
with such technicalities in a rigourous manner.
\subsection{Early time evolution}
The subsequent evolution, for larger values of $y$, can be studied by
expressing $\hh$ of equation~\eref{start} in terms of the invariant creation and
annihilation operators $\bid$ and $\bi$~\cite{lewis,acg,xi0}
and expand it in powers of $x \equiv a\, \hub/k \ll 1$ (see the condition~\eref{cc1}),
\beq
\hh = \frac{k}{a} 
\lqu \bid\,\bi + \frac{1}{2} 
+ \frac{i}{2}\,x \lt  \bi^2-(\bid)^2 \rt \rqu
+ O(x^2)
\ .
\label{Hosc}
\eeq
Upon neglecting the term of order $x$, the invariant operators
determine a basis $\{\ket{n}\}$ of orthonormal solutions to
\beq
i\, \ptau \kcs = \hh \kcs
\ ,
\eeq
such that (again, for $\tau\sim\tau_{\rm asy}$)
\beq
\hh\ket{n}\simeq
\frac{k}{a} \lt n+\frac{1}{2} \rt
\ket{n}
\equiv
\omega_n\,\ket{n}
\ ,
\label{kn}
\eeq 
for which quantum gravitational fluctuations are still negligible
\beq 
\Delta\hat O\ket{n}=\mathcal{O}(x^2)
\ .
\eeq
In this approximation, the complete (normalised) solution to our problem in regime~I
will then be given by
\beq
\kcs\simeq
\mathcal{N}^{1/2}\,
\sum_n e^{-i\,\omega_n\,(\tau-\tau_{\rm asy})}\ket{n}\pro{n}{\chi_{\rm asy}}
\ ,
\label{normed}
\eeq
where $\tau\gtrsim \tau_{\rm asy}\to-\infty$ and $\mathcal{N}=\mathcal{N}(\kappa)$
is a normalisation factor.
This form, which involves the usual states $\ket{n}$ with fixed number of quanta,
will now allow us to give a more transparent interpretation of the initial states
$\ket{\chi_{\rm asy}}$ in terms of physically meaningful quantities.
\par
We shall first obtain a relation between the asymptotic eigenvalue $\kappa$
and quantities that appear in the spectrum of the Hamiltonian~\eref{Hosc}.
For this purpose, it appears natural to equate the eigenvalue of the asymptotic
Hamiltonian~\eref {Hasy} on the asymptotic states $\ket{\kappa}$ with
the expectation value of $\hat H$ on the states in~\eref {normed}
for $\tau\to\tau_{\rm asy}$.
In particular,
\beq
\fl
\pro{n}{\chi_{\rm asy}}
&\simeq&
\left(\frac{3}{\pi^3}\right)^{1/4}
\sqrt{\frac{\sqrt{k}\,a\,\lp}{2^{n+1}\,n!}}
\int\limits_{-\infty}^{+\infty} \ud\phi\,
e^{i\,\sqrt{3}\,\kappa\,\lp\,\phi}
e^{-\frac{k}{2}\,a^2\,\phi^2}
H_n\left(\sqrt{k}\,a\,\phi\right)
\nonumber
\\
\fl
&=&
\frac{i^n}{\sqrt{2^{n+1}\,n!}}
\left(\frac{3}{\pi^3}\right)^{1/4}
\sqrt{\frac{\lp}{\sqrt{k}\,a}}\,e^{-3\,\lp^2\,\kappa^2/2\,k\,a^2}
\,
H_n\left(\sqrt{3}\,\frac{\lp\,\kappa}{\sqrt{k}\,a}\right)
\ ,
\label{dist}
\eeq
since the weighted Hermite polynomials are eigenfunctions of the Fourier transform.
This expression must be evaluated in the asymptotic regime by
assuming $a\sim k\,\lp$, thus yielding
\beq
\pro{n}{\chi_{\rm asy}}
\simeq
i^n
\left(\frac{3}{\pi^3}\right)^{1/4}
\frac{e^{-3\,\kappa^2/2\,k^3}}{\sqrt{2^{n+1}\,n!\,k^{3/2}}}
\,
H_n\left(\sqrt{3}\,\kappa\,k^{-3/2}\right)
\ .
\eeq
Since each state $\ket{n}$ of the mode $k$ contributes an amount of energy
equal to $\omega_n^{(k)}\simeq n\,k/a\simeq n/\lp$~\footnote{We are subtracting
the zero-mode energy.},
the energy (in units of the Hubble mass $\mathcal{H}$) stored in $\ket{\chi_{\rm asy}^{(k)}}$
for $k$ and $\kappa$ fixed is given by
\beq
\frac{E^{(k)}}{\mathcal{H}}
\simeq
\frac{\mathcal{N}^{(k)}}{2\,\lp\,\mathcal{H}}\,
\sqrt{\frac{3}{\pi^3}}\,
\frac{e^{-3\,\kappa^2/k^3}}{k^{3/2}}\,
\sum_{n=1}^\infty\,n\,R_n^{(k)}(\kappa)
\ ,
\label{Easy}
\eeq
where
\beq
R_n^{(k)}(\kappa)=
\left|\frac{\pro{n}{\chi_{\rm asy}}}{\pro{0}{\chi_{\rm asy}}}\right|^2
=\frac{H_n^2\left({\sqrt{3}\,\kappa}\,{k^{-3/2}}\right)}{2^{n}\,n!}
\ .
\eeq
We finally equate $E^{(k)}$ to the asymptotic energy
\beq
E_{\rm asy}^{(k)}=\frac{3\,\lp^2\,\kappa^2}{2\,a^3}
\simeq
\frac{3\,\kappa^2}{2\,\lp\,k^3}
\ ,
\label{E_asy}
\eeq
so that equation~\eref{Easy} determines the normalisation $\mathcal{N}^{(k)}$
as a function of $k$ and $E_{\rm asy}^{(k)}$,
\beq
\mathcal{N}^{(k)}
\simeq
2\,\sqrt{\frac{\pi^3}{3}}\,k^{3/2}\,
e^{2\,\lp\,E_{\rm asy}^{(k)}}
\,
\lp\,E_{\rm asy}^{(k)}
\left[
\sum_{n=1}^\infty\,n\,R_n^{(k)}
\right]^{-1}
\ ,
\eeq
where
\beq
R_n^{(k)}=
\frac{H_n^2\left(\sqrt{2\,\lp\,E_{\rm asy}^{(k)}}\right)}{2^n\,n!}
\ .
\eeq
All the expressions are now functions of the wave mode $k$ and the 
asymptotic energy $E_{\rm asy}^{(k)}$ which remain as free parameters.
\begin{figure}[t!] 
\begin{center}
\raisebox{4.5cm}{$R_n$}  
\includegraphics[width=0.6\textwidth]{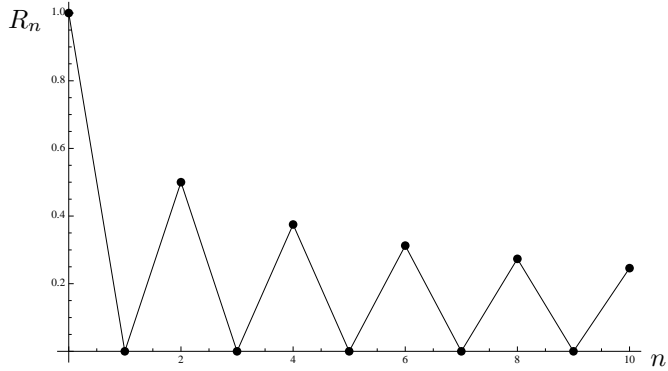}
$n$
\caption{Distribution in equation~\eref{Ralpha} for $n=0,\ldots,10$.
Only even excitations appear in the asymptotic states.}
\label{distri}
\end{center}
\end{figure}
\par
We recall that $k$ is constrained by equation~\eref{cc1} and
the total matter energy must be small with respect to $\mathcal{H}$,
otherwise the scale factor would not evolve {\em a la\/} de~Sitter.
Assuming the range of available $k$ extends over at least a few orders
of magnitude, the total energy
\beq
E_{\rm asy}=\int_{a\,\mathcal{H}}^{(a\,\mathcal{H})^{-1/2}}
\!\!
E_{\rm asy}^{(k)}\,\ud k
\ ,
\label{Etot}
\eeq
can be small only if~\footnote{See equation~\eref{grav2} below and Ref.~\cite{inprog}.} 
\beq
E_{\rm asy}^{(k)}
\ll
\frac{a^3\,\mathcal{H}}{2\,\lp^2}
\sim
k^3\,\lp^2\,\hub^2
\ ,
\label{cme}
\eeq
for all the allowed $k$.
The relative probability of finding $n$ excitations of each mode $k$ is then
approximately given by
\beq
R_n^{(k)}
\simeq
\frac{H_n^2(0)}{2^n\,n!}
=\frac{2^n\,\pi}{n!\,\Gamma^2((1-n)/2)}
\ .
\label{Ralpha}
\eeq
which is shown in Fig.~\ref{distri}.
\par
Technically, since $\ket{\chi_{\rm asy}}$ is not normalisable, the series
in equation~\eref{Easy} would diverge and $\mathcal{N}^{(k)}$ would vanish unless
the asymptotic states were regularised.
As we mentioned, both problems could be in fact cured by using wave-packets
instead of the simple plane-waves~\eref{sol}.
However, any such regularization would only affect the distributions $R_n^{(k)}$
at given $k$ for $n$ large, making them decrease faster for increasing $n$ than
the case shown in equation~\eref{Ralpha}.
We can therefore conclude that the states $\ket{\chi_{\rm asy}^{(k)}}$ are in
general peaked on the vacuum $\ket{n=0}$ and contain very small amounts of
(even) excitations for all the modes $k$.
\subsection{Matter fluctuations in the gravitational equation}
The asymptotic matter states we found are consistent solutions only if we can show
that their contribution to the evolution of the scale factor
of the Universe is small compared to the driving cosmological constant
$3\,\hub^2$.
This is not just tantamount to the condition~\eref{cme} above, but also requires that
their quantum gravitational fluctuations in the Einstein-Hamilton-Jacobi
equation for $a=a(\tau)$ remain small~\cite{bv}.
\par
From Ref.~\cite{acg}, we know that the equation that determines the
evolution of the wave-function $\psi=\psi(a)$ for the cosmic scale factor is given by
\beq
\fl
\lt \frac{\lp^2}{2\,a}\,\para^2 + \frac{a^3\,\hub^2}{2\,\lp^2}
+\sum_k\,\bra{\chi_k}{\hh}\ket{\chi_k} \rt \psi
=-\frac{\lp^2}{2\,a}\,\sum_k\,\bra{\chi_k}{\para^2}\ket{\chi_k}\,
\psi
\ ,
\label{grav}
\eeq
where the right hand side represents the aforementioned fluctuations, the sum in
$k$ formally represents the integral in equation~\eref{Etot} and expectation values are taken
on asymptotic states in the Heisenberg representation~\cite{xi0},
\beq
\fl
\ket{\chi_k}
=
\exp\lt{i \strut\displaystyle\int^\tau\bra{\chi_{\rm s}}{\hh}\ket{\chi_{\rm s}}\,\ud t}\rt
\ket{\chi_{\rm s}}
\simeq
\exp\lt{i \strut\displaystyle\int^\tau E^{(k)}_{\rm asy}\, \ud t}\rt
\ket{\chi_{\rm asy}}
\ .
\eeq
From equations~\eref{sol} and~\eref{E_asy}, one therefore obtains
\beq
\para^2 \ket{\chi_k} \simeq
i\,\para \left[\lt  \frac{E^{(k)}_{\rm asy}}{a\,\hub}
-\frac{3\,\lp^2\,\kappa^2}{2\,a^4\,\hub}
\rt
\ket{\chi_k}
\right]
\simeq
0
\ .
\label{dchik}
\eeq
It then follows that the gravitational equation reduces to
\beq
\lt \frac{\lp^2}{2\,a}\,\para^2 + \frac{a^3\,\hub^2}{2\,\lp^2} +E_{\rm asy}
\rt
\psi
=0
\ ,
\label{grav2}
\eeq
which yields the WKB solution corresponding to the de~Sitter evolution~\eref{ds}
provided the condition~\eref{cme} on the total matter energy we discussed before
holds~\footnote{A more complete analysis of the coupled gravity-matter dynamics
is being performed~\cite{inprog}.}.
\subsection{Intermediate regime}
\begin{figure}[t!] 
\begin{center} 
\raisebox{4.5cm}{$\gamma$} 
\includegraphics[width=0.6\textwidth]{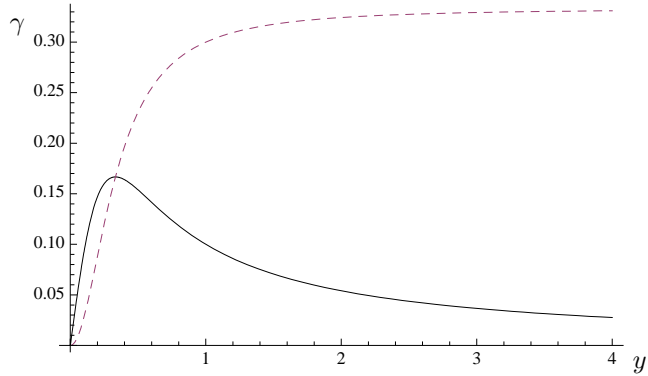}
$y$
\caption{Real (solid line) and imaginary (dashed line) parts of $\gamma(y)$.
Real part tends to zero for $y$ small (regime~II) and
large (regime~I) and has a maximum of $1/6$ at $y=1/3$.
Imaginary part vanishes at $y=0$ and tends to $1/3$
for large $y$.}
\label{fig:rhs}
\end{center}
\end{figure}
The results presented this far are quite interesting in
themselves but do not yet fully validate the findings of Ref.~\cite{xi0}.
In fact, from Ref.~\cite{xi0}, we already know that the effects on matter states
produced by gravitational fluctuations in regime~II ($y\ll 1$) are very small
(of order $\lp^2$), so we need to make sure that in between regimes~I and
II the fluctuations never become overwhelming with respect to the left hand
side in equation~\eref{start2}.
\par
We do not attempt at solving analytically equation~\eref{start2} for
arbitrary values of $y$, but we just note that there is no singularity
in the strength of the fluctuations at the transition between
regimes~I and II. 
Fig.~\ref{fig:rhs} shows the real and imaginary
parts of the coefficient that ``couples'' with the gravitational
fluctuations in~\eref{start2},
\beq
\gamma(y) \equiv \frac{y}{1-3\,i\,y}
\ .
\label{gamma}
\eeq
Looking at the graphs, it is evident that not only 
there is no singularity, but the real and imaginary parts
of $\gamma$ never grow larger than a few decimals.
Therefore, the gravitational perturbations never
become a dominant factor in the dynamics of the scalar field in de~Sitter
and the system reaches the regime~II analyzed in Ref.~\cite{xi0}
in a state that does not differ significantly from
(or even is just the same as) the one it was in during regime~I.
This clearly supports the results presented in Ref.~\cite{xi0}.
\section{Summary and conclusions}
\label{sec:summ}
We have investigated the influence of quantum gravitational fluctuations
on the evolution of a minimally coupled scalar field mode $k$ in a
spatially flat de~Sitter background for values of the scale factor near the
limit of validity of the semiclassical approximation, that is for $a \gtrsim k\,\lp$.
We have found that, contrary to na\"ive intuition, the effects of such
fluctuations do not grow indefinitely with increasing energy
({\em i.e.}, going backward in time) but they actually vanish and
the scalar field becomes asymptotically free in what we called regime~I.
This result is further supported by the fact that quantum gravitational fluctuations
have been shown to asymptotically vanish in the semiclassical equation that governs
the evolution of the cosmic scale factor~\cite{bv,acg,xi0}.
It is important to remark that these conclusions only apply
to those wave-numbers $k$ that satisfy the inequalities~\eref {cc1}.
Of course, this condition is not really restrictive in the flat de~Sitter
space we have considered here, since $a$ can be as small as possible.
However, (on general quantum mechanical considerations~\cite{gf})
one expects that the Universe was born into the semiclassical
regime with a non-zero initial scale factor and several approaches to quantum
cosmology further support a spatially closed initial geometry
(see, {\em e.g.}, Refs.~\cite{hs,loll}). 
The latter situation was actually taken into account in deriving equation~\eref{cc1},
which ensures the scalar field modes of relevance have a wavelength much
shorter than the Hubble length and are not affected by the spatial curvature
at very early times.
\par 
Of course, the possibility remains that $a\,\mathcal{H}\gtrsim 1$ initially,
which would rule out all values of $k$ with the asymptotic behaviour
found in Section~\ref{sec:asy}.
This occurs, for example, if the initial state of the Universe is exactly
the closed de~Sitter space-time of Ref.~\cite{hs}.
For such large initial values of $a$, the regime~I never happens and the
Universe is actually born in regime~II, in which quantum gravitational
fluctuations are always very small
(of order $\mathcal{H}^2\,\ell_{\rm P}^2$; see Ref.~\cite{xi0}).
Since we have also shown that between the early-time limit (if it exists)
and the perturbative late-time phase,
both the real and the imaginary part of the coupling
to gravitational fluctuations~\eref{gamma} always remain small,
we can conclude that the gravitational perturbations never become a
dominant factor for the dynamics of the scalar field in a de~Sitter Universe.
This argument validates and supports the result of Ref.~\cite{xi0}
where the effective dispersion relation and CMB spectrum
were computed.
\par
We wish to conclude by adding that,
even if regime~I never occurred in our actual Universe, the fact that a
matter-gravity system admits the kind of asymptotic freedom we have
found can be of theoretical interest for our understanding of gravitation.
In fact, although our approach is entirely different, we recalled in the
Introduction that a form of asymptotic freedom for pure gravity was
conjectured in Ref.~\cite{weinberg} and later supported, {\em e.g.\/},
in Refs.~\cite{souma}.
\ack
C.~A.~is supported by a Marco Polo fellowship of the University of Bologna and
wishes to thank the I.C.G., Portsmouth, UK, for their kind hospitality.
\section*{References}

\begin{thebibliography}{99}
%
\bibitem{ST}
M.B.~Green, J.H.~Schwarz and E.~Witten,
{\em Superstring Theory},
Cambridge University Press, Cambridge (1987);
M.~Gasperini and G.~Veneziano,
Phys.\ Rept.\  {\bf 373}, 1 (2003);
L.~McAllister and E.~Silverstein,
Gen.\ Rel.\ Grav.\  {\bf 40}, 565 (2008);
%
\bibitem{LQG}
M.~Bojowald,
Living Rev.\ Rel.\  {\bf 8}, 11 (2005);
A.~Ashtekar,
Nuovo Cim.\  {\bf 122B}, 135 (2007);
M.~Bojowald and R.~Tavakol,
``Loop Quantum Cosmology: Effective theories and oscillating Universes,''
arXiv:0802.4274.
%
\bibitem{birrell}
N.D.~Birrell and P.C.W.~Davies, {\em Quantum fields in curved space},
Cambridge University Press, Cambridge (1982);
I.~L.~Shapiro,
``Effective Action of Vacuum: Semiclassical Approach,''
arXiv:0801.0216.
%
\bibitem{hawking}
S.W. Hawking, Nature {\bf 248}, 30 (1974);
Comm. Math. Phys. {\bf 43}, 199 (1975).         
%
\bibitem{branden}
J.~Martin and R.H.~Brandenberger, Phys. Rev. D {\bf 63}, 123501
(2001).
%
\bibitem{dani}
U.H.~Danielsson,
Phys.\ Rev.\  D {\bf 66}, 023511 (2002).
%
\bibitem{tp}
%
D.~Lopez~Nacir and F.D.~Mazzitelli,
Phys.\ Rev.\  D {\bf 76}, 024013 (2007);
%
J.~Martin and C.~Ringeval,
JCAP {\bf 0608}, 009 (2006);
%
H.~Collins and R.~Holman,
Phys. Rev. D {\bf 74}, 045009 (2006);
%
B.~Greene, M.~Parikh and J.P.~van der Schaar,
JHEP {\bf 0604}, 057 (2006);
%
A.~Ashoorioon, J.L.~Hovdebo and R.B.~Mann,
Nucl. Phys. B {\bf 727}, 63 (2005);
%
L.~Sriramkumar and T.~Padmanabhan,
Phys. Rev. D {\bf 71}, 103512 (2005);
%
J.~de Boer, V.~Jejjala and D.~Minic,
Phys. Rev. D {\bf 71}, 044013 (2005);
%
K.~Ke,
Int. J. Mod. Phys. A {\bf 20}, 4331 (2005);
%
Q.G.~Huang and M.~Li,
Nucl. Phys. B {\bf 713}, 219 (2005);
%
S.~Hossenfelder,
Mod. Phys. Lett. A {\bf 19}, 2727 (2004);
%
M.~Porrati,
Phys.\ Lett.\  B {\bf 596}, 306 (2004);
%
G.~Calcagni,
Phys. Rev. D {\bf 70}, 103525 (2004);
%
S.~Shankaranarayanan and L.~Sriramkumar,
Phys. Rev. D {\bf 70}, 123520 (2004);
%
R.G.~Cai,
Phys. Lett. B {\bf 593}, 1 (2004);
%
M.~Porrati,
Phys. Lett. B {\bf 596}, 306 (2004);
%
M.~Fukuma, Y.~Kono and A.~Miwa,
Nucl. Phys. B {\bf 703}, 293 (2004);
%
S.~Hannestad,
JCAP {\bf 0404}, 002 (2004);
%
S.~Tsujikawa, P.~Singh and R.~Maartens,
Class. Quant. Grav.  {\bf 21}, 5767 (2004);
%
J.P.~van der Schaar,
JHEP {\bf 0401}, 070 (2004);
%
G.L.~Alberghi, K.~Goldstein and D.A.~Lowe,
Phys. Lett. B {\bf 578}, 247 (2004);
%
M.~Fukuma, Y.~Kono and A.~Miwa,
Nucl. Phys. B {\bf 682}, 377 (2004);
%
G.L.~Alberghi, R.~Casadio and A.~Tronconi,
Phys. Lett. B {\bf 579} (2004) 1.
%
V.~Bozza, M.~Giovannini and G.~Veneziano,
JCAP {\bf 0305}, 001 (2003);
%
S.~Tsujikawa, R.~Maartens and R.~Brandenberger,
Phys. Lett.  B {\bf 574}, 141 (2003);
%
O.~Elgaroy and S.~Hannestad,
Phys. Rev. D {\bf 68}, 123513 (2003);
%
E.~Keski-Vakkuri and M.S.~Sloth,
JCAP {\bf 0308}, 001 (2003);
%
M.B.~Einhorn and F.~Larsen,
Phys. Rev. D {\bf 68}, 064002 (2003);
S.~Cremonini,
Phys. Rev. D {\bf 68}, 063514 (2003);
%
A.A.~Starobinsky,
JETP Lett.  {\bf 73}, 371 (2001).
%
\bibitem{lim}
C.~Armendariz-Picon and E.A.~Lim,
JCAP {\bf 0312}, 006 (2003).
%
\bibitem{hs}
S.W.~Hawking,
``Volume Weighting in the No Boundary Proposal,''
arXiv:0710.2029;
J.B.~Hartle, S.W.~Hawking and T.~Hertog,
``The Classical Universes of the No-Boundary Quantum State,''
arXiv:0803.1663.
%
\bibitem{loll}
J.~Ambjorn, A.~Gorlich, J.~Jurkiewicz and R.~Loll,
Phys.\ Rev.\ Lett.\  {\bf 100}, 091304 (2008).
%
\bibitem{gf}
R.~Casadio,
Int.\ J.\ Mod.\ Phys.\  D {\bf 9}, 511 (2000).
%
\bibitem{WDW}
B.S.~DeWitt, Phys. Rev. {\bf 160}, 1113 (1967);
J.A.~Wheeler, in {\em Batelle rencontres: 1967 lectures in
mathematics and physics} edited by C.~DeWitt and J.A.~Wheeler
(Benjamin, New York, 1968).
%
\bibitem{bv}
R.~Brout and G.~Venturi, Phys. Rev. D {\bf 39}, 2436 (1989).
%
\bibitem{fvv}
F.~Finelli, G.~P.~Vacca and G.~Venturi,
Phys. Rev.  D {\bf 58}, 103514 (1998).
%
\bibitem{acg}
G.L.~Alberghi, R.~Casadio and A.~Gruppuso,
Phys. Rev. D {\bf 61},  084009 (2000).
%
\bibitem{xi0}
G.L.~Alberghi, R.~Casadio and A.~Tronconi,
Phys.\ Rev.\  D {\bf 74}, 103501 (2006);
G.L.~Alberghi, C.~Appignani, R.~Casadio, F.~Sbis\`a and A.~Tronconi,
Phys. Rev. D {\bf 77}, 044002 (2008)
%
\bibitem{kiefer}
C.~Kiefer and B.~Sandhoefer,
``Quantum Cosmology,''
arXiv:0804.0672.
%
\bibitem{weinberg}
S.~Weinberg, ``Ultraviolet divergences in quantum theories of gravitation,''
in {\em General relativity: an Einstein centenary survey},
edited by S.~Hawking and W.~Israel,
Cambridge University Press (Cambridge, 1979).
%
\bibitem{souma}
M.~Reuter,
Phys. Rev. D {\bf 57}, 971 (1998);
W.~Souma,
Prog.\ Theor.\ Phys.\  {\bf 102}, 181 (1999);
M.~Reuter and H.~Weyer,
``Background Independence and Asymptotic Safety in Conformally Reduced
Gravity,''
arXiv: 0801.3287.
%
\bibitem{lewis}
H.R.~Lewis and W.B.~Riesenfeld, J. Math. Phys. {\bf 10}, 1458
(1969);
X.-C.~Gao, J.-B.~Xu and T.-Z.~Qian, Phys. Rev. A {\bf 44}, 7016
(1991).
%
\bibitem{inprog}
C.~Appignani and R.~Casadio, work in progress.
%
\end{thebibliography}
\end{document}